\documentclass[preprint, 5p, twocolumn]{elsarticle}

\usepackage{lineno,hyperref}
\usepackage{lineno}
\usepackage{graphicx}
\usepackage{subcaption}
\usepackage{amsmath}
\usepackage{verbatim}
\usepackage{amssymb}
\usepackage{color}
\usepackage{graphicx}
\usepackage{caption}

\usepackage{mathrsfs}
\modulolinenumbers[5]

\journal{Journal of \LaTeX\ Templates}

\usepackage{orcidlink}
\newcommand{\rh}{r_{\text{h}}}

\begin{document}

\begin{frontmatter}

\title{$D$-dimensional black holes in extended Gauss-Bonnet gravity}


\author[address1,address2]{Jia-Zhou Liu\orcidlink{0009-0005-5430-4258}}
\author[address1,address2]{Si-Jiang Yang\orcidlink{0000-0002-8179-9365}}
\author[address1,address2]{Chun-Chun Zhu}
\author[address1,address2]{Yu-Xiao Liu\orcidlink{0000-0002-4117-4176}\corref{mycorrespondingauthor}}
\cortext[mycorrespondingauthor]{Corresponding author}
\ead{liuyx@lzu.edu.cn}
\address[address1]{Lanzhou Center for Theoretical Physics, Key Laboratory of Theoretical Physics of Gansu Province, Key Laboratory of Quantum Theory and Applications of MoE, Gansu Provincial Research Center for Basic Disciplines of Quantum Physics, Lanzhou University, Lanzhou 730000, China}
\address[address2]{Institute of Theoretical Physics $\&$ Research Center of Gravitation, School of Physical Science and Technology, Lanzhou University, Lanzhou 730000, China}

\begin{abstract}
	In this work, we present    charged spherically symmetric  anti-de Sitter black hole solutions in $d$ dimensions  ($d \geq 5$) within the framework of extended Gauss–Bonnet gravity.    Moreover, by employing dimensional regularization of the extended Gauss–Bonnet term, we further obtain charged anti-de Sitter black hole solutions in three- and four-dimensional spacetimes. Furthermore, we also obtain a three-dimensional rotating black hole solution in a special case. These solutions contain an independent vector-sector parameter and exhibit different horizon structures compared with the standard charged Gauss--Bonnet branch. For the charged solutions, we also give a concise Iyer-Wald thermodynamic analysis.
	Nonetheless, their asymptotic behavior at spatial infinity remains consistent with that of the corresponding black holes in general relativity.
\end{abstract}


\end{frontmatter}



\section{Introduction}\label{sec:intro}

	Lovelock theory of gravity~\cite{Lovelock:1971yv} represents a prominent higher-curvature generalization of general relativity. It extends the Einstein-Hilbert action by incorporating higher-order curvature invariants, while preserving second-order field equations and avoiding the introduction of additional propagating degrees of freedom. In $d$ dimensions, the Lovelock action is given by a linear combination of the first $\lceil d/2 \rceil$ Lovelock invariants, since higher-order terms either vanish identically or reduce to topological contributions. These higher-curvature invariants do not affect the gravitational field equations in spacetimes with $d < 5$, where their variation yields either vanishing or purely topological contributions.

It is well known that  Gauss–Bonnet gravity represents the simplest nontrivial case within the Lovelock class, whose invariant contributions routinely occur
in the low-energy limit of quantum theories of gravity, such as string theory~\cite{Zwiebach:1985uq,Nepomechie:1985us,Ferrara:1996hh,Antoniadis:1997eg,Gross:1986mw,Candelas:1985en}.  In higher dimensions, however, the Gauss-Bonnet term becomes dynamically relevant, as it is no longer a total derivative but contributes nontrivially to the gravitational field equations through its quadratic dependence on curvature.  The general static spherically symmetric black hole solutions in higher dimensional Gauss-Bonnet gravity were found by Boulware et al.~\cite{Boulware:1985wk}, and the properties of these black holes were also studied~\cite{Wheeler:1985qd,Myers:1988ze,Jacobson:1993xs,Cai:2001dz,Cai:2003kt,Garraffo:2008hu}.  Gauss–Bonnet gravity lives on the Riemannian sector of the theory, meaning that both torsion and nonmetricity should be zero.

In recent years, there has been increasing scholarly interest in exploring Gauss–Bonnet gravity within the framework of Weyl geometry~\cite{Bahamonde:2025qtc,BeltranJimenez:2014iie,TerenteDiaz:2023kgc,Charmousis:2025jpx,Haghani:2014zra,BeltranJimenez:2015pnp,Lutfuoglu:2025ldc,Eichhorn:2025pgy},. In particular, Ref.~\cite{Bahamonde:2025qtc} presents an exact five-dimensional black hole solution in extended Gauss–Bonnet gravity, while Ref.\cite{Charmousis:2025jpx} obtains  exact four-dimensional black hole solutions by applying dimensional regularization to the Gauss–Bonnet term, thereby constructing a novel theory. In  Weyl geometry , the connection is symmetric but is no longer metric-compatible. Instead, the non-metricity is encoded by a vector field $W_\mu$, giving rise to a Weyl connection. In this context, an extended formulation of Gauss–Bonnet gravity has been proposed in arbitrary dimension, with the action constructed from curvature invariants associated with the Weyl connection~\cite{BeltranJimenez:2014iie}. Remarkably, such a construction can naturally lead to vector–tensor theories with Gauss–Bonnet term ~\cite{Charmousis:2025jpx,Haghani:2014zra,Will:1972zz,Heisenberg:2014rta,Heisenberg:2017mzp}.

The present work builds on the exact five-dimensional solution of Ref.~\cite{Bahamonde:2025qtc}. We extend this construction to arbitrary dimension and include a Maxwell charge. The resulting family contains the known five-dimensional neutral solution as the special case $d=5$ and $Q=0$, while the standard Einstein-Maxwell-Gauss-Bonnet-AdS and RN-AdS limits are recovered when the vector-sector parameter is switched off. By considering dimensional regularization of the extended Gauss-Bonnet term, we also obtain the charged AdS black hole solutions in three- and four-dimensional spacetime. Furthermore, we present a three-dimensional rotating black hole solution. For the charged solutions, we include a concise Iyer-Wald analysis of the vector-sector contribution to the first law.

The outline of the paper is as follows. Section~\ref{SS2} provides a review of extended Gauss-Bonnet gravity in Weyl geometry. In Sec.~\ref{SS3}, we present the higher-dimensional  charged AdS black hole solution in the extended Gauss-Bonnet gravity. In Sec.~\ref{SS4}, by considering dimensional regularization of the extended Gauss-Bonnet term, we give the charged AdS black hole solutions in three- or four-dimensional spacetime. In Sec.~\ref{SS5}, building upon the results of Sec.~\ref{SS4}, we present a three-dimensional rotating black hole solution.
Finally, Sec.~\ref{SS6} presents a summary and discussion of the work.

\section{EXTENDED GAUSS-BONNET GRAVITY IN WEYL GEOMETRY }  \label{SS2}


Two years after Einstein formulated general relativity, Weyl introduced the concept of Weyl symmetry~\cite{Weyl:1918ib,Hehl:1994ue,Romero:2012hs}. He proposed that a proper generalization of flat spacetime to curved spacetime should incorporate not only local Lorentz transformations, but also conformal  transformations of vectors parallel transported around closed loops. This idea led to the development of Weyl geometry, an extension of Riemannian geometry that incorporates an additional vector field $W_\lambda$ via~
\begin{equation}
	\tilde{\nabla}_{\lambda}g_{\mu\nu}=-2g_{\mu\nu}W_{\lambda}.
\end{equation}
The components of the Weyl connection  are given by
\begin{equation}
	\tilde{\Gamma}^\lambda_{\mu\nu} = \Gamma^\lambda_{\mu\nu} + \delta^\lambda_\mu W_\nu + \delta^\lambda_\nu W_\mu - g_{\mu\nu} W^\lambda,
\end{equation}
where $\Gamma^\lambda_{\mu\nu}$ is the  Christoffel symbols  constructed from the metric $g_{\mu\nu}$. The Riemann tensor is defined as
\begin{eqnarray}
	\tilde{R}^\lambda{}_{\rho\mu\nu}&=&\partial_\mu\tilde{\Gamma}^\lambda{}_{\rho\nu}-\partial_\nu\tilde{\Gamma}^\lambda{}_{\rho\mu}+\tilde{\Gamma}^\lambda{}_{\sigma\mu}\tilde{\Gamma}^\sigma{}_{\rho\nu}-\tilde{\Gamma}^\lambda{}_{\sigma\nu}\tilde{\Gamma}^\sigma{}_{\rho\mu}
	\nonumber \\
	&=&R^{\lambda}{}_{\rho\mu\nu}-g_{\rho[\mu}\delta^{\lambda}{}_{\nu]}W_{\alpha}W^{\alpha}-W_{[\mu}g_{\nu]\rho}W^{\lambda}\nonumber \\
	&-&\delta^{\lambda}{}_{[\mu}W_{\nu]}W_{\rho}-\nabla_{[\mu}W^{\lambda}g_{\nu]\rho}+2\delta^{\lambda}{}_{\rho}\nabla_{[\mu}W_{\nu]}\nonumber \\&+&2\nabla_{[\mu}W_{|\rho|}\delta^{\lambda}{}_{\nu]},
\end{eqnarray}
where $R^{\lambda}{}_{\rho\mu\nu}$ denotes the Riemann tensor built solely   from the  Christoffel symbols.
The corresponding Ricci and co-Ricci tensors are defined as
\begin{equation}\tilde{R}_{\mu\nu}=\tilde{R}^{\lambda}{}_{\mu\lambda\nu},\quad\hat{R}_{\mu\nu}=\tilde{R}_{\mu}{}^{\lambda}{}_{\nu\lambda}.\end{equation}
The generalized Gauss--Bonnet invariant in Weyl geometry is given by~\cite{BeltranJimenez:2014iie,Bahamonde:2025qtc}
\begin{equation}\tilde{\mathcal{G}}=\tilde{R}_{\mu\nu\lambda\rho}\tilde{R}^{\lambda\rho\mu\nu}-(\tilde{R}_{\mu\nu}+\hat{R}_{\mu\nu})(\tilde{R}^{\nu\mu}+\hat{R}^{\nu\mu})+\tilde{R}^2.\end{equation}
Moreover, the Gauss–Bonnet term constructed from the Weyl connection can be recast into the form~\cite{BeltranJimenez:2014iie,Charmousis:2025jpx}
\begin{equation}
	\begin{aligned}
		\tilde{\mathcal{G}} &= \mathcal{G} + (d-3)(d-4)\left(4G^{\mu\nu}W_\mu W_\nu\right) \\
		&\quad + (d-4)(d-3)(d-2) \cdot 4W^2\nabla_\mu W^\mu \\
		&\quad + (d-4)(d-3)(d-2)(d-1)W^4 \\
		&\quad + (d-3)\nabla_\mu J^\mu,
	\end{aligned}
\end{equation}

where $\mathcal{G} = R^{2} - 4 R_{\mu\nu} R^{\mu\nu} + R_{\mu\nu\alpha\beta} R^{\mu\nu\alpha\beta}$ is the standard Gauss–Bonnet term constructed from the Riemannian curvature. We define $W^2 \equiv W_\mu W^\mu$ and $W^4 \equiv (W^2)^2$, and introduce the vector current
\begin{equation}
	J^\mu  =8G^{\mu\nu}W_\nu+4(d-2)\left[W^\mu(W^2+\nabla_\nu W^\nu)-W_\nu\nabla^\nu W^\mu\right].
\end{equation}
By ignoring the boundary term $\nabla_\mu J^\mu$, the extra term coming from the Weyl part is given by
\begin{equation}
	\begin{aligned}
		\mathcal{G}_{w} &= \tilde{\mathcal{G}} - \mathcal{G} \\
		&= (d-4)(d-3)\Bigl(4G^{\mu\nu}W_{\mu}W_{\nu} \\
		&\quad + (d-2)\bigl(4W^{2}\nabla_{\mu}W^{\mu} + (d-1)W^{4}\bigr)\Bigr).
	\end{aligned}
\end{equation}
In four dimensions, the term $\mathcal{G}_w$ becomes topological in nature, reducing to a boundary term that carries no dynamical content in the field equations.
It follows that the Gauss–Bonnet  invariant constructed from the Weyl connection can be expressed as the standard Gauss–Bonnet term computed with the Levi-Civita connection, plus a total derivative and an additional term that vanishes identically in four dimensions.

We now consider a generalized action for extended Gauss–Bonnet gravity in $d$-dimensional spacetime:
\begin{equation}S=\frac{1}{16\pi G}\int d^dx\sqrt{-g}\left(R-2\Lambda+\alpha\mathcal{G}+\beta\mathcal{G}_{w}-F^{\mu\nu}F_{\mu\nu}\right), \label{S}\end{equation}
where $\Lambda$ denotes the cosmological constant,  $\alpha$ and $\beta$ represent the  coupling constants, and the electromagnetic tensor $F_{\mu\nu}$ related to the electromagnetic field is
\begin{eqnarray}
	F_{\mu\nu}=\partial_{\mu}A_\nu-\partial_{\nu}A_{\mu} .
	\label{Fa1}
\end{eqnarray}

When $\alpha=\beta$, the theory represents a natural extension of Gauss–Bonnet gravity within Weyl geometry, where both curvature and nonmetricity play essential roles due to the non-vanishing of the latter.

When $\alpha \neq \beta$, and if the Weyl vector field is interpreted as an ordinary vector field rather than a geometric object encoding non-metricity, the action \eqref{S} can be reformulated within the standard framework of Riemannian geometry. In this case, it corresponds to a vector–tensor theory of gravity~\cite{Charmousis:2025jpx,Haghani:2014zra,Domenech:2018vqj,Moffat:2005si} augmented by the Gauss–Bonnet term. Therefore, this reformulation allows the theory to be interpreted as a standard higher-curvature modification of General Relativity coupled to a dynamical vector field, rather than as a purely geometric extension rooted in Weyl non-metricity.

The gravitational field equations in the framework of the extended Gauss-Bonnet gravity  can be derived by varying the action \eqref{S} with respect to the metric tensor $g^{\mu\nu}$:
\begin{equation}
	G_{\mu\nu}+\Lambda g_{\mu\nu}=T^\mathrm{GB}_{\mu\nu}+  T^\mathrm{W}_{\mu\nu}+ T^\mathrm{F}_{\mu\nu} ,
	\label{modified}
\end{equation}
where
\begin{equation}
	\begin{aligned}
		T^\mathrm{GB}_{\mu\nu} &= \alpha\bigl(\mathcal{G} - 2RR_{\mu\nu} + 4R_{\mu\gamma}R^{\gamma}_{\ \nu} \\
		&\quad + 4R_{\gamma\delta}R^{\gamma\ \delta}_{\ \mu\ \nu} - 2R_{\mu\gamma\delta\lambda}R_{\nu}^{\ \gamma\delta\lambda}\bigr),
	\end{aligned}
\end{equation}

\begin{equation}
	\begin{aligned}
		T_{\mu\nu}^\mathrm{W} &= 2\beta(d-4)(d-3)\Bigl[W^{\alpha}W_{\alpha}G_{\mu\nu} + W_{\mu}W_{\nu}R \\
		&\quad + W^{\alpha}W^{\beta}R_{\alpha\beta}g_{\mu\nu} - 2W_{\mu}W^{\alpha}R_{\alpha\nu} - 2W_{\nu}W^{\alpha}R_{\alpha\mu} \\
		&\quad + \nabla_{\alpha}\nabla_{\mu}\left(W^{\alpha}W_{\nu}\right) + \nabla_{\alpha}\nabla_{\nu}\left(W^{\alpha}W_{\mu}\right) - \nabla^{2}\left(W_{\mu}W_{\nu}\right) \\
		&\quad - g_{\mu\nu}\nabla_{\alpha}\nabla_{\beta}\left(W^{\alpha}W^{\beta}\right) - \nabla_{\mu}\nabla_{\nu}(W^{\alpha}W_{\alpha}) \\
		&\quad + g_{\mu\nu}\nabla^{2}(W^{\alpha}W_{\alpha})\Bigr] \\
		&\quad - \beta(d-4)(d-3)(d-2)(d-1)\left[2W^{2}W_{\mu}W_{\nu} - \frac{1}{2}g_{\mu\nu}W^4\right] \\
		&\quad + 4\beta(d-4)(d-3)(d-2)\Bigl[(\nabla_\mu W_\alpha)W_\nu W^\alpha \\
		&\quad + (\nabla_\nu W_\alpha)W_\mu W^\alpha - W_{\mu}W_{\nu}\nabla_{\alpha}W^{\alpha} \\
		&\quad - W^{\alpha}W^{\beta}(\nabla_{\beta}W_{\alpha})\Bigr],
		\label{WW}
	\end{aligned}
\end{equation}
and
\begin{equation}
	T^\mathrm{F}_{\mu\nu}=2F_{\mu\alpha}F_{\nu}^{\alpha}-\frac{1}{2}g_{\mu\nu}F^{\alpha\beta}F_{\alpha\beta}.
	\label{TF}
\end{equation}
Similarly, by varying the action \eqref{S} with respect to the Weyl vector $W_\mu$ we can obtain the equations of motion for the corresponding  field:
\begin{multline}
	2G_{\mu\alpha}W^{\alpha} + (d-2)(d-1)W_{\alpha}W^{\alpha}W_{\mu} \\
	- 2(d-2)\left(W_{\mu}\nabla_{\alpha}W^{\alpha} - W_{\alpha}\nabla_{\mu}W^{\alpha}\right) = 0.
	\label{WG}
\end{multline}
Having derived the field equations, we proceed in the following section to present charged AdS black hole solutions in  $d$-dimensional spacetime ($d \geq 5$).

\section{CHARGED BLACK HOLE SOLUTIONS IN $d$-DIMENSIONAL SPACETIME} \label{SS3}
We consider the following metric ansatz for the $d-$dimensional  static spherically symmetric spacetime:
\begin{equation}ds^2=-A(r)dt^2+S(r)dr^2+r^2d\Omega_{d-2}^2. \label{mm}\end{equation}
The Weyl vector field  and the $U(1)$ vector field with static spherically symmetry are given by
\begin{eqnarray}
	&&W_\mu\mathrm{d}x^\mu=w_0(r)\mathrm{d}t+w_1(r)\mathrm{d}r \label{tr}, \\
	&&  A_\mu\mathrm{d}x^\mu=\Phi(r)\mathrm{d}t 	\label{Fa}.
\end{eqnarray}
Since there is no gauge symmetry associated with the vector field $W_\mu$, in general $w_1$ cannot be set to zero.

For the convenience of readers who wish to reproduce the equations of motion and check the solutions, we also give the reduced effective action in generic dimension. Substituting the ansatz~\eqref{mm},~\eqref{tr}, and~\eqref{Fa} into the $d$-dimensional action~\eqref{S} and integrating over the angular coordinates, we obtain
\begin{equation}
	S_{\rm eff}
	=\frac{\Omega_{d-2}}{16\pi G_d}\int dt\,dr\,\mathcal{L}_{\rm eff}.
	\label{Seffd}
\end{equation}
The corresponding effective Lagrangian is
\begin{equation}
\begin{aligned}
	\mathcal{L}_{\rm eff}
	=&(d-2)\sqrt{A(r)S(r)}
	\frac{d}{dr}\Bigg[
	r^{d-3}\left(1-\frac{1}{S(r)}\right) \\
	&\quad
	+\alpha(d-3)(d-4)r^{d-5}
	\left(1-\frac{1}{S(r)}\right)^2
	\Bigg] \\
	&+2\Lambda r^{d-2}\sqrt{A(r)S(r)}
	+\frac{2r^{d-2}}{\sqrt{A(r)S(r)}}\Phi'(r)^2 \\
	&+\beta(d-3)(d-2)(d-4)
	r^{d-2}\sqrt{A(r)S(r)}\,\mathcal{I}_{W},
\end{aligned}
	\label{Leffd}
\end{equation}
where
\begin{equation}
\begin{aligned}
	\mathcal{I}_{W}
	=&\frac{2\left[rS'(r)+(d-3)S(r)(S(r)-1)\right]}
	{r^2A(r)S(r)^2}w_0(r)^2 \\
	&+\frac{2\left[rA'(r)-(d-3)A(r)(S(r)-1)\right]}
	{r^2A(r)S(r)^2}w_1(r)^2 \\
	&+4\left(\frac{w_1(r)^2}{S(r)}
	-\frac{w_0(r)^2}{A(r)}\right)
	\frac{1}{r^{d-2}\sqrt{A(r)S(r)}} \\
	&\quad \times
	\frac{d}{dr}\left[
	r^{d-2}\sqrt{\frac{A(r)}{S(r)}}w_1(r)
	\right] \\
	&+(d-1)\left(\frac{w_1(r)^2}{S(r)}
	-\frac{w_0(r)^2}{A(r)}\right)^2 .
\end{aligned}
	\label{IWd}
\end{equation}
Here the prime denotes the derivative with respect to $r$. Varying the reduced action~\eqref{Seffd} with respect to $A(r)$, $S(r)$, $w_0(r)$, $w_1(r)$, and $\Phi(r)$ gives the radial equations.

\subsection{ Case A: $\alpha=\beta$ }
First, we consider the scenario involving  $\alpha=\beta$. This theory constitutes a natural extension of Gauss–Bonnet gravity within Weyl geometry, where not only curvature is present but nonmetricity is also non-vanishing.
By combining the ansatzs~\eqref{mm},~\eqref{tr} and the field equations~\eqref{modified} - \eqref{WG}, we can obtain the static spherically symmetric charged black hole solution in  $d-$dimensional  spacetime with $d\geq 5$:
\begin{eqnarray}
	A(r)&=&1+\frac{\frac{r^2}{l^2}-\frac{m}{r^{d-3}}
	-\frac{(d-3)(d-4)\alpha \eta^2}{4r^{2d-4}}
	+\frac{2(d-3)Q^2}{(d-2)r^{2(d-3)}}}
	{1-\frac{(d-3)(d-4)\alpha \eta}{r^{d-1}}},\\ 
	S(r)&=&\frac{1}{A(r)},\\
	w_0(r)&=&\frac{- 2r^{d-3}A(r)+2r^{d-3}+ \eta}{4 r^{d-2}},\label{Q11}\\
	w_1(r)&=&\frac{w_0(r)}{A(r)},\label{Q12}\\
	\Phi(r)&=& -\frac{Q}{r^{d-3}}.
\end{eqnarray}

Here, $l$ denotes the AdS radius, which is related to the cosmological constant via $\Lambda=-(d-1)(d-2)/(2l^2)$, and $\eta$ is an integration constant. It is evident that the obtained solution deviates significantly from the charged AdS black hole solutions in Gauss–Bonnet gravity~\cite{Cvetic:2001bk}. For the solutions obtained below, the relation $A(r)S(r)=1$ follows from the field equations and is not imposed as an additional assumption. In the special case where $d=5$ and $Q=0$, the solution coincides with that reported in Ref.~\cite{Bahamonde:2025qtc}. The parameter $\eta$ is not a redefinition of $M$, $Q$, or $l$: it appears in powers of $r$ different from the mass, charge, and cosmological terms, and it also fixes the independent vector-field profile. We therefore interpret $\eta$ as an independent vector hair parameter associated with the vector sector.

Furthermore, when either $\alpha = \beta = 0$ or $\eta = 0$, the solution reduces to the standard $d$-dimensional Reissner–Nordström(RN)–AdS black hole. One notices that the constant $\eta$  appears in the zero-component of    Weyl vector as $\propto \eta/ r^{d-2}$. This constant may be interpreted as characterizing a novel dipole-type configuration, potentially related to dilations~\cite{Bahamonde:2025qtc}.
In the asymptotic limit $r \rightarrow \infty$, the metric function behaves as $A(r) \rightarrow 1 + r^2/l^2$, indicating that the spacetime is asymptotically AdS.
Compared with the usual charged Einstein-Maxwell-Gauss-Bonnet-AdS branch, the additional $\eta$-dependent terms can change the zeros of the metric function and, in suitable regions of parameter space, generate an additional horizon.
Another notable feature of the solution is the presence of a spacetime singularity, which can be studied through the Kretschmann scalar  
$K$, derived from the Riemann tensor:
\begin{eqnarray}
	K&=&R_{\alpha\beta\delta\gamma}R^{\alpha\beta\delta\gamma}.
\end{eqnarray}
When  $\alpha \eta > 0$, we have
\begin{equation}
	K\propto \left\{
	\begin{aligned}
		\frac{1}{r^{4(d-2)}}	&,&\quad   r  \to 0;\\
		\frac{1}{\left(1-\frac{(d-3)(d-4)\alpha \eta}{r^{d-1}}\right)^6}	&,&\quad r \to \left(\alpha \eta (d - 4)(d - 3) \right)^{\frac{1}{d - 1}};	\\
		\frac{2(d-2)(d-3)}{l^4}&,&\quad   r  \to \infty.
	\end{aligned}	\label{Dr1}
	\right. 
\end{equation}
When $\alpha \eta \leq 0$, we have
\begin{equation}
	K\propto \left\{
	\begin{aligned}
		\frac{1}{r^{4(d-2)}}&,&\quad   r  \to 0;\\
		\frac{2(d-2)(d-3)}{l^4}&,&\quad  r  \to \infty.
	\end{aligned}	\label{Dr2}
	\right.
\end{equation}
\begin{figure}[htbp]
	\centering
	\begin{subfigure}[b]{0.28\textwidth}
		\includegraphics[width=\linewidth]{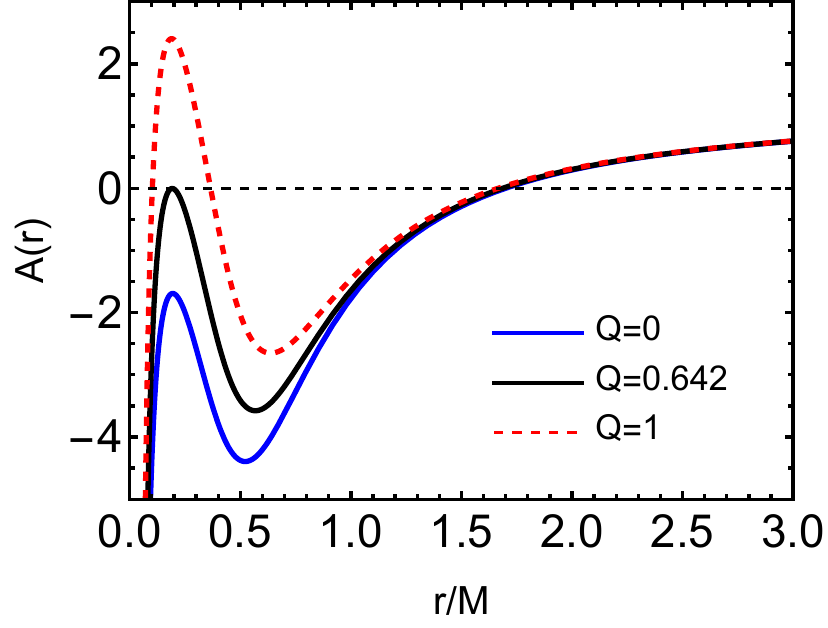}
		\caption{}
		\label{p1}
	\end{subfigure}\hfill
	\begin{subfigure}[b]{0.28\textwidth}
		\includegraphics[width=\linewidth]{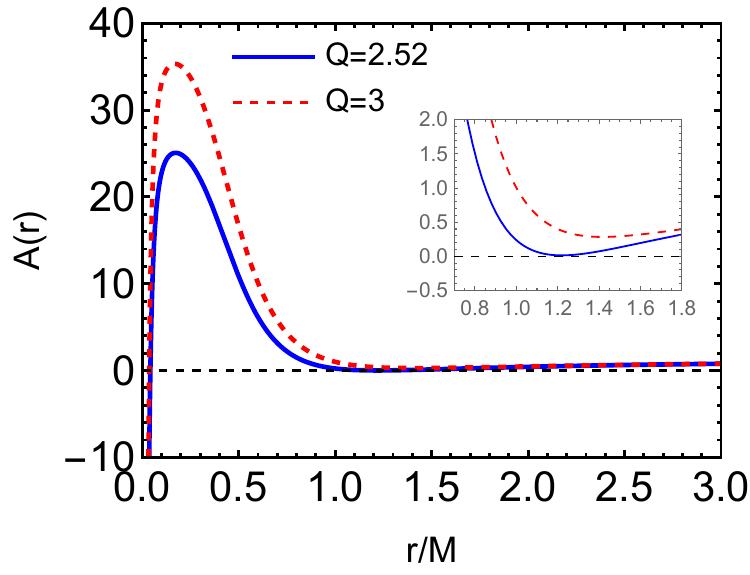}
		\caption{}
		\label{p2}
	\end{subfigure}
	\caption{The metric function $A(r)$ for the black hole with parameters $d = 5$, $M = 1$, $1/l^2 = 0.01$, $\alpha = 0.2$, and $\eta = -0.2$, is shown for different values of the charge parameter $Q$.}
	\label{A(r)}
\end{figure}
From this result, we observe that the spacetime exhibits a singularity only at $r = 0$ when $\alpha \eta \leq 0$. However, for $\alpha \eta>0$,  singularities arise at both $r = 0$ and $r = \left(\alpha \eta (d - 4)(d - 3) \right)^{1/(d - 1)}$. Similarly, when  $\alpha \eta>0$, the Ricci scalar $R$ and the  contraction
of two Ricci tensors $R_{\mu\nu} R^{\mu\nu}$ also diverge at $r = \left(\alpha \eta (d - 4)(d - 3) \right)^{1/(d - 1)}$. In certain parameter space, a singularity may not be shielded by event horizon, potentially leading to the formation of naked singularity. Since the case $\alpha \eta > 0$ may lead to the appearance of a non-zero naked singularity, we shall focus on the physically acceptable scenario with $\alpha \eta \leq 0$ in the subsequent analysis.
We set
\begin{align}
	B(r) =A(r)(1 - \frac{(d-3)(d-4)\alpha \eta}{r^{d-1}}).\label{eq:B} 
\end{align}
Since $\alpha \eta \leq 0$, the zeros of the function $A(r)$ coincide with those of $B(r)$ in this case.

Next, we examine the asymptotic behavior of the metric function $B(r)$. In the limit $r \to 0$, it approaches 
\[
B(r) \to -\frac{(d-3)(d-4)\alpha \eta^2}{4r^{2d-4}}.
\]
While for $r \to \infty$, it tends to $B(r) \to r^2 / l^2$. When $\alpha < 0$ and $\eta > 0$, we find that $B(r) > 0$ in both the $r \to 0$ and $r \to \infty$ limits. In this case, by analyzing the condition $B'(r) = 0$ and taking into account the parameter constraints $M > 0$, $Q^2 > 0$, and $l^2 > 0$, we find that there exists an extremum of $B(r)$ at some finite radius $r > 0$. Combined with the fact that $B(r) > 0$ in both the $r \to 0$ and $r \to \infty$ limits, this indicates that the metric function $B(r)$ must develop a minimum in the positive $r$ region. Consequently, the spacetime admits a structure similar to that of the RN-AdS black hole, featuring the possibility of two horizons.

On the other hand, for $\alpha \geq 0$ and $\eta \leq 0$, we find $B(r) < 0$ as $r \to 0$ and $B(r) > 0$ as $r \to \infty$. In this case, by analyzing the condition $B'(r) = 0$, we find that the function $B(r)$ admits two extrema. Combined with the asymptotic behavior $B(r) < 0$ as $r \to 0$ and $B(r) > 0$ as $r \to \infty$, it follows that the black hole solution may admit one, two, or even three horizons, depending on the specific values of the parameters, as shown in Figs.~\ref{p1} and ~\ref{p2}.

\subsection{ Case B: $\alpha\neq\beta$ }
Next, we consider the scenario involving  $\alpha\neq\beta$. The action \eqref{S} corresponds to a vector–tensor theory of gravity augmented by a Gauss–Bonnet term  within the framework of Riemannian geometry.
Similarly, by combining the ansatz~\eqref{mm},~\eqref{tr} and the field equations~\eqref{modified}- \eqref{WG}, we can obtain the solution in the $d-$dimensional ($d \geq 5$) spacetime:
\begin{align}
	A(r) &= 1 - \frac{\beta \eta}{2(\alpha-\beta)r^{d-3}} + \frac{r^{2}}{2\gamma(\alpha-\beta)} \bigg(1 \nonumber \\
	&\quad + \epsilon \biggl[1 - \frac{4\gamma(\alpha-\beta)}{l^2} + \frac{\gamma(4(\alpha-\beta)m - 2\beta \eta)}{r^{d-1}} \nonumber \\
	&\quad + \frac{\gamma^2\alpha\beta \eta^2}{r^{2d-2}} - \frac{8(d-4)(d-3)^2(\alpha-\beta)Q^2}{(d-2)r^{2d-4}}\biggr]^{1/2} \bigg), \label{SB} \\
	S(r) &= \frac{1}{A(r)}, \label{a2} 
\end{align}
where $\gamma = (d - 3)(d - 4)$ and $\epsilon = \pm 1$, while $m$ and $\eta$ appear as integration constants. 
It is observed that the constant $\eta$ also appears in the Weyl vector with a scaling behavior proportional to $\eta/r^{d-2}$, and it also contributes a correction to the metric. In the special case of $d=5$ and $Q=0$, this solution coincides with that presented in Ref.~\cite{Bahamonde:2025qtc}.  The black hole solution obtained here resembles the charged AdS black hole solutions in Gauss–Bonnet gravity, but includes additional corrections involving the parameters $\beta$ and $\eta$. As in Gauss–Bonnet gravity, the present theory admits two distinct branches of solutions. When $\beta = 0$, the solution reduces to the $d$-dimensional   charged AdS black hole in Gauss–Bonnet gravity~\cite{Boulware:1985wk,Garraffo:2008hu}.
We note that in the limit $r \to 0$, the asymptotic behavior of the square root part in the metric function $A(r)$ behaves as $\gamma^2 \alpha \beta \eta^2/{r^{2d-2}}$. Therefore, when the product $\alpha \beta > 0$, the argument of the square root remains positive in the $r \to 0$ limit, ensuring the regularity of the metric function near the origin.

When $M=Q=\eta=0$, the vacuum solution in Eq.~\eqref{SB} is
\begin{eqnarray}
	A(r)=1+\frac{r^{2}}{2\gamma(\alpha-\beta)}\left(1   +\epsilon\sqrt{1-\frac{4\gamma(\alpha-\beta)}{l^2}}\right).
\end{eqnarray}
Since $\alpha - \beta > 0$, the condition $0 \leq 1 - 4\gamma(\alpha - \beta)/l^2$ must be satisfied to ensure the validity of the solution. Under this condition, the action \eqref{S} admits two AdS vacua with effective cosmological constants
\begin{equation}
	l_{\mathrm{eff}}^{2} = \frac{l^2}{2} \left( 1 -\epsilon \sqrt{1 - \frac{4\gamma(\alpha - \beta)}{l^2}} \right).
\end{equation}
If $\alpha - \beta < 0$, the spacetime remains AdS provided the $\epsilon = -1$ branch is selected.

Similar to the Gauss–Bonnet case, when $Q = 0$ and $1/l^2 = 0$, the solution in Eq.~\eqref{SB} becomes asymptotically Schwarzschild for the $\epsilon = -1$ branch. On the other hand, the $\epsilon = +1$ branch yields an asymptotically Schwarzschild-AdS solution with a negative gravitational mass, indicating an inherent instability. In the context of Gauss–Bonnet gravity, it was shown by Boulware and Deser~\cite{Boulware:1985wk} that the $\epsilon = +1$ branch leads to a ghost-like graviton and is thus unstable, whereas the $\epsilon = -1$ branch is stable and free of ghost excitations.
\begin{figure}[htbp]
	\centering
	\begin{subfigure}[b]{0.28\textwidth}
		\includegraphics[width=\linewidth]{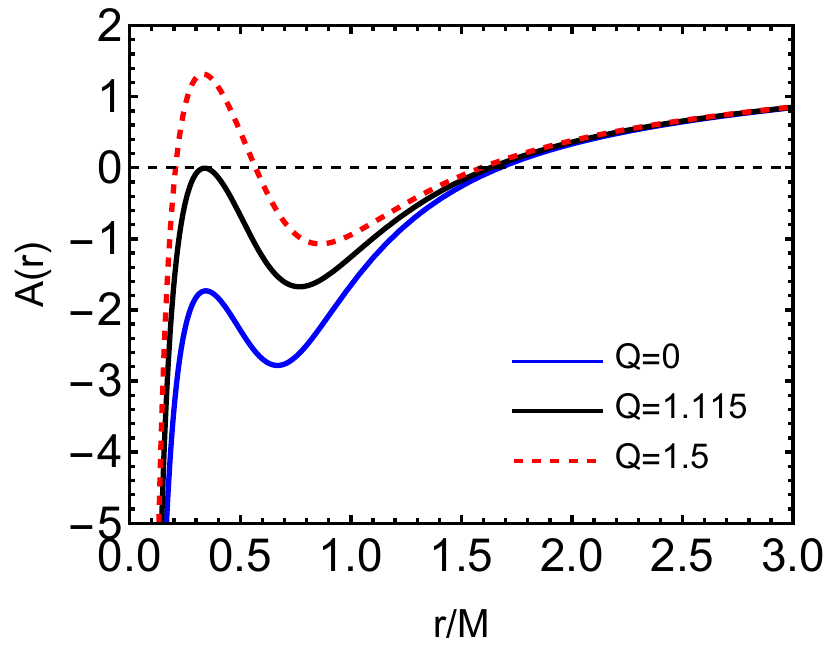}
		\caption{}
		\label{cp11}
	\end{subfigure}\hfill
	\begin{subfigure}[b]{0.28\textwidth}
		\includegraphics[width=\linewidth]{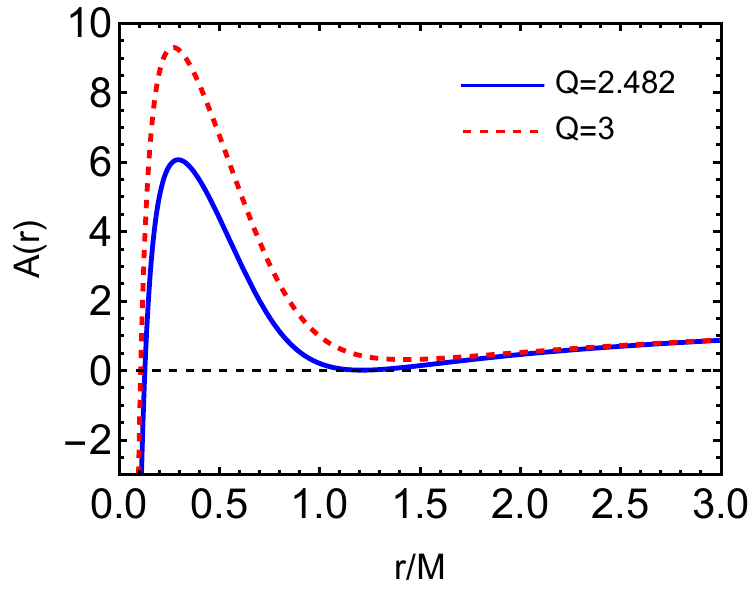}
		\caption{}
		\label{cp12}
	\end{subfigure}
	\caption{The metric function $A(r)$ for the black hole with parameters $d = 5$, $M = 1$, $1/l^2 = 0.01$, $\alpha = 0.18$, $\beta = 0.2$, $\eta = -0.61$, and for different values of the charge parameter $Q$.}
	\label{A1(r)}
\end{figure}

Furthermore, when $\alpha < \beta < 0$ and $\eta \beta \leq4(\alpha - \beta)m$, the metric function $A(r)$ remains real and well-defined for all $r \in (0, \infty)$, with no imaginary components. In this regime, the profile of $A(r)$ closely resembles that of the RN-AdS black hole.

In contrast, for $0 < \alpha < \beta$ and $\eta \beta \leq4(\alpha - \beta)m$, the metric function $A(r)$ also remains real and regular throughout the domain $r \in (0, \infty)$. In this case, the black hole exhibits a richer causal structure, potentially admitting one, two, or even three horizons depending on the choice of parameters, as shown in Figs. \ref{cp11} and \ref{cp12}.

\subsection{Thermodynamics of the charged black holes}
For the charged black-hole solutions obtained above, we perform a concise Iyer-Wald analysis~\cite{Wald:1993nt,Iyer:1994ys}. Denoting \(E_R^{abcd}\equiv\partial\mathcal{L}/\partial R_{abcd}\), the relevant presymplectic potential is
\begin{equation}
\begin{aligned}
	\mathbf{\Theta}_{a_2\cdots a_d}
	= \Big[&
	2{E_R}^{aefh}\nabla_h\delta g_{ef}
	-2(\nabla_h{E_R}^{aefh})\delta g_{ef}\\
	&+\frac{2\beta\mathcal C_d W^2}{\kappa}
	g^{ae}\delta W_e
	-\frac{2}{\kappa}F^{ae}\delta A_e
	\Big]\\
	&\times\varepsilon_{a a_2\cdots a_d},
\end{aligned}
\end{equation}
where \(\kappa=8\pi G_d\) and \(\mathcal C_d=(d-4)(d-3)(d-2)\). The corresponding Noether charge can be written as
\begin{equation}
\begin{aligned}
	(\mathbf{Q}_\xi)_{a_3\cdots a_d}
	= \Big[&
	-{E_R}^{abef}\nabla_e\xi_f
	-2\xi_e\nabla_f{E_R}^{abef}\\
	&-\frac{2\beta\mathcal C_d W^2}{\kappa}
	g^{ab}W_f\xi^f
	-\frac{1}{\kappa}F^{ab}A_e\xi^e
	\Big]\\
	&\times\varepsilon_{ab a_3\cdots a_d}.
\end{aligned}
\end{equation}
For the horizon Killing vector \(\xi_H\), the closed variation current gives
\begin{equation}
	\int_{S_r}\left(\delta\mathbf{Q}_{\xi_H}-\xi_H\cdot\mathbf{\Theta}\right)
	-\int_{S_h}\left(\delta\mathbf{Q}_{\xi_H}-\xi_H\cdot\mathbf{\Theta}\right)=0 .
\end{equation}
At infinity,
\begin{equation}
	\delta H_{\xi}
	=\int_{S_\infty}\left(\delta\mathbf{Q}_{\xi}
	-\xi\cdot\mathbf{\Theta}[\Phi,\delta\Phi]\right).
	\label{eq_deltaH}
\end{equation}
Evaluating this boundary term gives
\begin{equation}
\begin{aligned}
	\delta E_W&
	=\frac{(d-2)\Omega_{d-2}}{16\pi G_d}\,\delta m=\delta M_{\rm ADM},\\
	M_{\rm ADM}&=\frac{(d-2)\Omega_{d-2}}{16\pi G_d}\,m .
\end{aligned}
\end{equation}

 For the Case B charged solution, let \(\rh\) be the largest root of \(A(r)\). In the physical branch \(\epsilon=-1\), the horizon equation \(A(\rh)=0\) gives the mass parameter :
\begin{equation}
\begin{aligned}
	m=&\rh^{d-3}
	+\gamma(\alpha-\beta)\rh^{d-5}
	-\frac{\gamma\beta\eta}{\rh^2}
	-\frac{\gamma\beta\eta^2}{4\rh^{d-1}}\\
	&+\frac{2(d-3)Q^2}{(d-2)\rh^{d-3}},
\end{aligned}
\end{equation}
 The Hawking temperature is obtained directly from the metric function,
\begin{equation}
	T=\frac{1}{4\pi}\left.\frac{dA(r)}{dr}\right|_{r=\rh},
\end{equation}
where the derivative is taken at fixed \(m,Q,\eta\), before imposing \(A(\rh)=0\). The vector-sector term in \(-\xi_H\cdot\mathbf{\Theta}\) contributes at the horizon, so the binormal Wald density alone,
\begin{equation}
	\begin{aligned}
		S_W
		=-2\pi\int_{S_h}{E_R}^{abcd}\epsilon_{ab}\epsilon_{cd}
		\\
		=\frac{\Omega_{d-2}\rh^{d-2}}{4G_d}
		\left(1+\frac{2(d-2)(d-3)\alpha}{\rh^2}\right),
	\end{aligned}
\end{equation}
is not the thermodynamic entropy $S_{\rm th}$. The full horizon variation then gives
\begin{equation}
	\int_{S_h}\left(\delta\mathbf{Q}_{\xi_H}-\xi_H\cdot\mathbf{\Theta}\right)-\Phi_Q\delta Q-\Phi_\eta\delta\eta=T\delta S_{\rm th},
\end{equation}
\begin{equation}
\begin{aligned}
	S_{\rm th}
	=\frac{\Omega_{d-2}}{4G_d}\Big[
	&\rh^{d-2}
	+2(d-2)(d-3)(\alpha-\beta)\rh^{d-4}\\
	&+\frac{(d-2)(d-3)(d-4)\beta\eta}{\rh}
	\Big],
\end{aligned}
\end{equation}
with zero integration constant. Treating \(\eta\) as a thermodynamic variable, the black-hole first law is obtained as
\begin{equation}
	\delta M_{\rm ADM}
	=T\delta S_{\rm th}+\Phi_Q\delta Q+\Phi_\eta\delta\eta,
\end{equation}
where
\begin{equation}
	\Phi_Q=\frac{(d-3)\Omega_{d-2}Q}{4\pi G_d\rh^{d-3}},
\end{equation}
and
\begin{equation}
\begin{aligned}
	\Phi_\eta=-\frac{(d-2)\Omega_{d-2}(d-3)(d-4)\beta}{16\pi G_d}
	\left(\frac{1}{\rh^2}+\frac{\eta}{2\rh^{d-1}}+\frac{4\pi T}{\rh}\right),
\end{aligned}
\end{equation}
with the theory parameters $\alpha$, $\beta$ kept fixed.

	\section{CHARGED BLACK HOLE SOLUTIONS IN THREE- AND FOUR- DIMENSIONAL SPACETIME} \label{SS4}

By treating the spacetime dimension $d$ as a continuous parameter of the theory and appropriately rescaling the Gauss–Bonnet coupling constant, it is possible to construct well-defined versions of the theory in $d = 4$ and even $d = 3$~\cite{Glavan:2019inb,Fernandes:2022zrq,Hennigar:2020lsl,Gurses:2020ofy,Bonifacio:2020vbk,Mahapatra:2020rds,Konoplya:2020qqh}. This approach represents a generalization of a method that was applied some time ago to obtain the $d \to 2$ limit of general relativity~\cite{Mann:1992ar}, and is consistent with a dimensional reduction
procedure~\cite{Lu:2020iav,Kobayashi:2020wqy}.

Here the target dimension is fixed to \(p=3\) or \(p=4\) before taking the regularized limit.

The resulting theory takes the form of a scalar–tensor theory of gravity, characterized by a non-minimal coupling between a scalar field and the curvature terms inherited from the dimensionally continued Gauss–Bonnet term~\cite{Lu:2020iav,Kobayashi:2020wqy,Fernandes:2021dsb}.
In this 
construction, the Gauss–Bonnet coupling constant is 
redefined as $\alpha \to \alpha/(d - p)$, followed by taking the limit $d \to p$.
The following action is the scalar--tensor effective action obtained from the regularized Gauss--Bonnet sector for the target dimensions \(p=3\) and \(p=4\):
\begin{equation}
	\begin{aligned}
		S &= \frac{1}{16\pi G} 
		\int d^{p}x \, \sqrt{-g} \Big[ 
		R - 2\Lambda
		\\
		&\quad
		+ \alpha \big(
		\phi \, \mathcal{G}
		+ 4 G^{ab} \nabla_{a}\phi \, \nabla_{b}\phi
		- 4 (\nabla\phi)^{2} \square\phi
		+ 2 \big[ (\nabla\phi)^{2} \big]^{2}
		\big)
		\Big] .
	\end{aligned}
	\label{eq:action}
\end{equation}

Here, $\phi$ is associated with the regularization procedure, which involves a conformal transformation between the metrics. Specifically, the conformally related metrics are given by
\begin{equation}
	\hat{g}_{\mu\nu} = e^{2\phi} g_{\mu\nu}.
\end{equation}
In analogy with the dimensional regularization approach applied to the Gauss--Bonnet term, we apply the same procedure to the Gauss--Bonnet invariant constructed from the Weyl connection, aiming to define a consistent low-dimensional extension of the theory. For the low-dimensional targets considered here, \(p=3\) and \(p=4\), we rescale \(\beta\to\beta/(d-p)\) and take the limit \(d\to p\). The regularized Weyl-sector contribution can then be written in the compact form
\begin{equation}
\begin{aligned}
	\mathcal{L}_{W}^{(p)}
	&\equiv \lim_{d\to p}\frac{\mathcal{G}_{w}(d)}{d-p}  \\
	&= \lim_{d\to p}
	\frac{(d-4)(d-3)}{d-p}
	\Bigl[
	4G^{\mu\nu}W_{\mu}W_{\nu}  \\
	&\quad
	+ (d-2)\bigl(
	4W^{2}\nabla_{\mu}W^{\mu}
	+(d-1)W^{4}
	\bigr)
	\Bigr].
\end{aligned}
\label{LWp}
\end{equation}
Combining this term with the scalar--tensor part from the regularized Gauss--Bonnet term gives the low-dimensional effective action used below. The generic \(d\)-dimensional parent action is already given in Sec.~2, while the expressions here are the \(p=3\) and \(p=4\) specializations:
\begin{equation}
\begin{aligned}
	S_p
	= \frac{1}{16\pi G}\int d^p x\,\sqrt{-g}\,
	\Bigl[
	R-2\Lambda
	+\alpha\mathcal{L}_{\phi}
	+\beta\mathcal{L}_{W}^{(p)}
	-F^{\mu\nu}F_{\mu\nu}
	\Bigr],
\end{aligned}
\label{Sp}
\end{equation}
where
\begin{equation}
\begin{aligned}
	\mathcal{L}_{\phi}
	= \phi\mathcal{G}
	+4G^{ab}\nabla_a\phi\nabla_b\phi
	-4(\nabla\phi)^2\square\phi
	+2\big[(\nabla\phi)^2\big]^2 .
\end{aligned}
\label{Lphi}
\end{equation}

For later use, we write explicitly the two specializations of Eq.~\eqref{Sp}. For \(p=3\), one obtains
\begin{equation}
	\begin{aligned}
		S = &\, \frac{1}{16\pi G} \int d^{3}x \, \sqrt{-g} \Big[
		R - 2\Lambda
		+ \alpha \big(
		\phi \, \mathcal{G}
		+ 4 G^{ab} \partial_{a}\phi \, \partial_{b}\phi
		\\
		& \quad
		- 4 (\partial\phi)^{2} \square\phi
		+ 2 \big[ (\nabla\phi)^{2} \big]^{2}
		\big)
		- \beta \big(
		4 G^{\mu\nu} W_{\mu} W_{\nu}
		\\
		& \quad
		+ 4 W^{2} \nabla_{\mu} W^{\mu}
		+ 2 W^{4}
		\big)
		- F^{\mu\nu} F_{\mu\nu}
		\Big] ,
	\end{aligned}
	\label{S3}
\end{equation}
For \(p=4\), the corresponding action becomes
\begin{equation}
	\begin{aligned}
		S &= \frac{1}{16\pi G}\int d^{4}x\sqrt{-g}\Bigl[R - 2\Lambda + \alpha\Bigl(\phi\mathcal{G} + 4G^{ab}\partial_{a}\phi\partial_{b}\phi \\
		&\quad - 4(\partial\phi)^{2}\square\phi + 2((\nabla\phi)^{2})^{2}\Bigr) + \beta\Bigl(4G^{\mu\nu}W_{\mu}W_{\nu} \\
		&\quad + \bigl(8W^{2}\nabla_{\mu}W^{\mu} + 6W^{4}\bigr)\Bigr) - F^{\mu\nu}F_{\mu\nu}\Bigr],
		\label{S4}
	\end{aligned}
\end{equation}
Having obtained the black hole solution in $d$ dimensions in the previous section, we are now motivated to adopt a similar ansatz in the subsequent analysis:
\begin{eqnarray}
	ds^2&=&-A(r)dt^2+\frac{1}{A(r)h(r)^2}dr^2+r^2d\Omega_{d-2}^2,\nonumber\\
	W_\mu\mathrm{d}x^\mu&=&w_0(r)\mathrm{d}t+\frac{w_0(r)}{A(r)h(r)}\mathrm{d}r .
	\label{34a}
\end{eqnarray}
\subsection{ Case A: $p=3$ }
Upon substituting the ansatz  \eqref{34a} and \eqref{Fa} to  the action \eqref{S3}, we obtain the effective Lagrangian 
density for variables $(A,h,\phi,w_0,\Phi)$:
\begin{equation}
	\begin{aligned}
		\mathcal{L}_{\mathrm{eff}} &= -\frac{1}{h(r)}\biggl[2r\Lambda + h(r)h^{\prime}(r)\Big(2A(r) - 4\beta w_0^{2}(r) \\
		&\quad + rA^{\prime}(r)\Big) + 4r\alpha A(r)^{2}h(r)^{3}h^{\prime}(r)\phi^{\prime}(r)^{3} \\
		&\quad + h(r)^{2}\Big(2A^{\prime}(r) - r\big(2\Phi^{\prime}(r)^{2} - A^{\prime\prime}(r)\big)\Big) \\
		&\quad - 2\alpha A(r)h(r)^{4}\phi^{\prime}(r)^{2}\Big(A^{\prime}(r)\big(1 - 2r\phi^{\prime}(r)\big) \\
		&\quad + A(r)\big(-2\phi^{\prime}(r) + r\phi^{\prime}(r)^{2} - 2r\phi^{\prime\prime}(r)\big)\Big)\biggr].
	\end{aligned}
\end{equation}
The condition $h(r)=1$ used below is obtained from the field equations for this solution branch and is not an additional assumption.
Then the equation coming from $\delta A$ reads
\begin{equation}4\alpha A(r)\phi^{\prime}(r)(-1+r\phi^{\prime}(r))\left(\phi^{\prime}(r)^2+\phi^{\prime\prime}(r)\right)=0. \end{equation}
The equation admits the following solution:
\begin{equation}
	\phi(r)=\log(r/C),
\end{equation}
where $C$ is an integration constant. The equation coming from $\delta \phi$ is then identically satisfied, while that from $\delta h$ yields
\begin{equation}
	\begin{aligned}
		&\frac{2\alpha A(r) \left(r A'(r) - A(r)\right)}{r^3} + A'(r) - 2r \Phi(r)^2 \\
		&+ 2\Lambda r - 8\beta w_0(r) w_0'(r) = 0.
		\label{f1}
	\end{aligned}
\end{equation}
By considering the equations of motion for the electromagnetic field $\frac{d}{dr}[r\Phi'(r)]=0$ and the Weyl field~\eqref{WG}, we obtain the following result:
\begin{eqnarray}
	\Phi(r)&=&-Q \log(r/r_0), \label{q1}\\
	w_0(r)&=&\frac{- A(r)+ \eta}{2 r} \label{w2}.
\end{eqnarray}
By substituting Eqs.~\eqref{q1} and \eqref{w2} into Eq.~\eqref{f1}, we obtain
\begin{multline}
	2\beta\eta^2 + \left(2\beta\eta r + r^3\right) A'(r) \\
	+ 2A(r) \left(r(\alpha - \beta) A'(r) - 2\beta\eta\right) \\
	+ 2(\beta - \alpha) A(r)^2 - 2Q^2 r^2 + 2\Lambda r^4 = 0.
\end{multline}
The corresponding black hole solutions depend on the spacetime dimension and the relation between the coupling constants $\alpha$ and $\beta$. The metric function $A(r)$ takes different forms depending on the relation between the coupling constants. When $\alpha = \beta$, we have:\begin{eqnarray}
	A(r)&=\frac{-M+\frac{\alpha \eta^2}{r^{2}}+\frac{1}{l^2}r^2-2Q^2\log(r/r_0)}{1+\frac{2\alpha \eta}{r^{2}}} .
	\label{S22}
\end{eqnarray}When $\alpha \ne \beta$, the solution reads as:
\begin{equation}
	\begin{aligned}
		A(r) &= -\frac{\beta \eta}{\alpha-\beta} - \frac{r^{2}}{2(\alpha-\beta)} \bigg(1 \\
		&\quad + \epsilon \biggl[1 + \frac{4(\alpha-\beta)}{l^2} - \frac{8(\alpha-\beta)Q^2\log(r/r_0)}{r^{2}} \\
		&\quad - \frac{4(\alpha-\beta)M + 4\beta \eta}{r^{2}} + \frac{4\alpha\beta \eta^2}{r^{4}}\biggr]^{1/2} \bigg).
	\end{aligned}
	\label{S23}
\end{equation}
In an independent and contemporaneous study~\cite{Alkac:2025zzi}, the same three-dimensional black hole solutions were derived.
To recover the three-dimensional black hole solution from the $d \to 3$ limit of the higher-dimensional case, one must implement the transformations $\alpha \to -\alpha$ and $\beta \to -\beta$ .
For the scenario  $\alpha=\beta$, when $\alpha = 0$ or $\eta = 0$, the resulting black hole solution reduces to the standard three-dimensional charged BTZ black hole solution \cite{Banados:1992wn}. Similarly, when $\alpha \neq \beta$ and in particular for $\alpha = 0$, the black hole solution in Eq.~\eqref{S23} corresponds to a solution within the framework of vector–tensor gravity.
\subsection{ Case B: $p=4$ }
Similarly, by applying the same procedure, the equations of motion can be 
derived from the effective Lagrangian density associated with the action \eqref{S4}:
\begin{align}
	&\alpha + r^3 A'(r) + A(r) \left(-2\alpha - 2\alpha r A'(r) + r^2\right) + 2\alpha r A'(r) \nonumber \\
	&\quad + \alpha A(r)^2 + \Lambda r^4 + r^4 \Phi'(r)^2 + 8\beta r^3 w_0(r) w_0'(r) \nonumber \\
	&\quad + 4\beta r^2 w_0(r)^2 - r^2 = 0, \\
	&4\alpha \bigl(A(r)(1 - r\phi'(r))^2 - 1\bigr)\left(\phi'(r)^2 + \phi''(r)\right) = 0.
\end{align}
Finally, by considering the equations of motion for the electromagnetic field $\frac{d}{dr}[r^2\Phi'(r)]=0$ and the Weyl field~\eqref{WG}, we can obtain the black hole solution in four dimensions.
For $\alpha = \beta$:
\begin{eqnarray}
	A(r) &=&1+\frac{\frac{1}{l^2}r^2-\frac{2M}{r}+\frac{Q^2}{r^{2}}-\frac{\alpha \eta^2}{4r^{4}}}{1-\frac{\alpha \eta}{r^{3}}},\\
	w_0(r)&=&\frac{- 2r A(r)+2r +\eta}{4 r^2},\\
	w_1(r)&=&\frac{	w_0(r)}{A(r)},\\
	\Phi(r)&=&-\frac{Q}{r}.
\end{eqnarray}
For  $\alpha \ne \beta$:
\begin{align}
	A(r) &= 1 - \frac{\beta \eta}{2(\alpha-\beta)r} + \frac{r^{2}}{2(\alpha-\beta)} \bigg(1 \nonumber \\
	&\quad + \epsilon \biggl[1 + \frac{\alpha\beta \eta^2}{r^{6}} - \frac{4(\alpha-\beta)}{l^2} - \frac{4(\alpha-\beta)Q^2}{r^{4}} \nonumber \\
	&\quad + \frac{8(\alpha-\beta)M - 2\beta \eta}{r^{3}}\biggr]^{1/2} \bigg) \label{S42} .
\end{align}
 The black hole solution exhibits similarities to those found in lower-dimensional Gauss–Bonnet gravity~\cite{Konoplya:2020ibi,Fernandes:2020rpa,Hennigar:2020drx}. Similar to the $d$-dimensional black hole discussed in Sec.~\ref{SS2}, the four-dimensional black hole can possess one, two, or even three horizons, depending on the  parameter space.  Similarly, when $\alpha \neq \beta$ and in particular for $\alpha = 0$, the black hole solution in Eq.~\eqref{S42} corresponds to a solution within the framework of vector–tensor gravity. In the case $\Lambda = 0$, the solutions coincide with those reported in Ref.~\cite{Charmousis:2025jpx}.

\section{ROTATING  BLACK HOLE SOLUTION IN THREE- DIMENSIONAL SPACETIME } \label{SS5}

In this section, we present a three-dimensional rotating black hole solution. When $\alpha = 0$, the action \eqref{S3} reduces to  vector–tensor theory of gravity. 
The stationary axial symmetric black hole metric in the three-dimensional spacetime has the form
\begin{equation}
	{d}{s}^{2}=-A(r) {dt}^{2}+S(r
	){dr}^{2}+r^{2}[K(r) dt+{d}\varphi]^{2}.
	\label{g3}
\end{equation}
We consider the ansatz for the vector field in three dimensions as given in Eq.~\eqref{tr}.

By varying the action with respect to the metric and the vector field, we obtain the corresponding field equations.
\begin{equation}
	G_{\mu\nu}+\Lambda g_{\mu\nu}=  T^{W}_{\mu\nu},
	\label{modified2}
\end{equation}
\begin{equation}
	\begin{aligned}
		2G_{\mu\alpha}W^{\alpha} &+ (d-2)(d-1)W_{\alpha}W^{\alpha}W_{\mu} \\
		&- 2(d-2)\bigl(W_{\mu}\nabla_{\alpha}W^{\alpha} - W_{\alpha}\nabla_{\mu}W^{\alpha}\bigr) = 0,
	\end{aligned}
	\label{FF2}
\end{equation}
By substituting Eqs.~\eqref{g3} and \eqref{tr} into filed equations,  we finally obtain the corresponding black hole solution:
\begin{align}
	A(r) &= \eta + \frac{r^{2}}{2\beta}\left(1 + \epsilon\sqrt{1 - \frac{4\beta}{l^2} + \frac{4\beta (\eta + M)}{r^{2}}}\right) + \frac{J^2}{4r^2}, \label{eq:A} \\
	S(r) &= \frac{1}{A(r)}, \label{eq:S} \\
	K(r) &= -\frac{J}{2r^2}, \label{eq:K} \\
	w_0(r) &= \frac{-A(r) + \frac{J^2}{4r^2} + \eta}{2r}, \label{eq:w0} \\
	w_1(r) &= \frac{w_0(r)}{A(r) - \frac{J^2}{4r^2}}. \label{eq:w1}
\end{align}
When $J = 0$, the black hole solution coincides with the three-dimensional black hole solution given in Eq.~\eqref{S23}, under the condition $\alpha = 0$ and $Q = 0$. Following the previous discussion, we choose $\epsilon = -1$.
When $\beta > 0$ and $\eta > -M$, or $\beta < 0$ and $\eta < -M$, the metric function $A(r)$ remains real and regular throughout the domain $r \in (0, \infty)$. Similar to the rotating BTZ black hole in three dimensions, it also possesses two horizons.

\section{ CONCLUSION AND DISCUSSION} \label{SS6}
In this work, we presented exact $d$-dimensional charged AdS black hole solutions in the extended Gauss–Bonnet gravity theory. When  $\alpha=\beta$, it is a natural extension of Gauss–Bonnet gravity in Weyl geometry, where both curvature and nonmetricity are present.
In this case,  the properties of the charged AdS black hole differ significantly from those of the charged AdS black hole in Gauss--Bonnet gravity. The additional integration constant $\eta$ is associated with the vector sector and cannot be absorbed into the mass, charge, or AdS radius. Depending on the  parameter space, the black hole can possess one, two, or even three horizons. Moreover, within specific regions of parameter space, the black hole solution can exhibit a curvature singularity at a nonzero radial coordinate. On the other hand, when $\alpha \neq \beta$, and if the Weyl vector field is interpreted as an ordinary vector field rather than a geometric object encoding non-metricity, the action can be recast within the framework of Riemannian geometry. In this case, it corresponds to a vector–tensor theory  augmented by a Gauss–Bonnet term. Although the resulting charged AdS black hole solution may appear similar to those in Gauss--Bonnet gravity, its physical properties still differ considerably. Depending on the parameter space, the black hole may possess one, two, or three horizons. The Iyer-Wald analysis gives the ADM energy and shows that the vector hair parameter enters the horizon thermodynamics through its conjugate term in the first law.

Furthermore, in three or four dimensions, the Gauss–Bonnet term is topological and therefore does not contribute dynamically to the field equations. However, by employing dimensional regularization of the extended Gauss–Bonnet term, we obtained charged AdS black hole solutions in both three- and four-dimensional spacetimes. By setting $\alpha = 0$, we  obtained a
three-dimensional rotating black hole solution.

While these black hole solutions differ from those in Gauss–Bonnet gravity, they exhibit the same asymptotic behavior at infinity as the corresponding black holes in general relativity.
\section*{Acknowledgments}
	This work was supported by the National Natural Science Foundation of China (Grants No. 12475056, No. 12247101, No.12305065 and No. 12475055 ), the China Postdoctoral Science Foundation (Grant No.2023M731468),
the 111 Project (Grant No. B20063), 
and Gansu Province's Top Leading Talent Support Plan.


\bibliography{ref}
\end{document}